\documentclass[amsmath,amssymb,aps,prd,nofootinbib,superscriptaddress,groupedaddress,twocolumn]{revtex4-2}
\usepackage[utf8]{inputenc}
\usepackage{mathrsfs}
\usepackage{bm}
\usepackage{url}
\usepackage[normalem]{ulem}
\usepackage{mathtools}
\usepackage{array}
\newcolumntype{P}[1]{>{\centering\arraybackslash}p{#1}}
\newcolumntype{M}[1]{>{\centering\arraybackslash}m{#1}}
\usepackage[caption=false]{subfig}
\usepackage{dcolumn}
\usepackage{graphicx, epsfig}
\usepackage[dvipsnames]{xcolor}
\usepackage{mathrsfs}
\usepackage{bm}
\usepackage{yhmath}
\usepackage[caption=false]{subfig}
\usepackage[normalem]{ulem}
\usepackage{mathtools}
\usepackage{bigints}
\usepackage{float}
\usepackage[colorlinks = true, linkcolor = purple, urlcolor  = blue, citecolor = blue, anchorcolor = blue]{hyperref}
\usepackage{float}
\usepackage{multirow}

\newcommand{\be}{\begin{equation}}
\newcommand{\ee}{\end{equation}}
\newcommand{\ba}{\begin{eqnarray}}
\newcommand{\ea}{\end{eqnarray}}

\newcommand{\nn}{\nonumber}

\def\lp4{$\lambda \phi^4$}

\begin{document}

\title{Resonance structures in kink-antikink scattering in a quantum vacuum
}
\author{Mainak Mukhopadhyay} \email{mkm7190@psu.edu}
\affiliation{Department of Physics; Department of Astronomy \& Astrophysics; Center for Multimessenger Astrophysics, Institute for Gravitation and the Cosmos, The Pennsylvania State University, University Park, PA 16802, USA.
}
\author{Tanmay Vachaspati} \email{tvachasp@asu.edu}
\affiliation{Physics Department, Arizona State University,
Tempe, AZ 85287, USA}

\date{\today}

\begin{abstract}
We investigate kink-antikink scattering in the $\lambda \phi^4$ model in the presence of an additional scalar field,
$\psi$, that is in its quantum vacuum and interacts with $\phi$ via a $\xi \phi^2\psi^2$ term where $\xi$ is
the coupling.
The final state of such a scattering is either a bound state with eventual annihilation or a reflection of the 
kink-antikink pair. Without the $\psi$ field, the outcome is known to depend fractally on the initial velocity 
of the kink-antikink pair. In the quantum vacuum of the $\psi$ field, the fractal dependence gets modified
and disappears above a critical interaction strength, $\xi \approx 0.1$.
\end{abstract}

\maketitle

\section{Introduction}
\label{sec:intro}
%

The interactions and scattering of topological defects has been an important area of 
research with applications to condensed matter systems~\cite{Zurek:1985qw,PhysRevLett.119.013902,PhysRevLett.105.075701,Chuang:1991zz,Bowick:1992rz,Hendry1994,Ruutu1996,Bauerle1996,PhysRevLett.89.080603,PhysRevLett.84.4966,PhysRevLett.91.197001,Beugnon2017} and cosmology~\cite{Zurek:1996sj,Vachaspati:1998vc,Kibble:2001wj,Planck:2013mgr}.
Since topological defects are solutions to classical equations of motion, most analyses
have considered classical scattering. This has led to important results such as
the universality of vortex reconnections~\cite{Vilenkin:2000jqa} and fractal behavior in the scattering of 
$\lambda\phi^4$ kinks in 
1+1 dimensions~\cite{kudryavtsev1975solitonlike,sugiyama1979kink,Campbell:1983xu,
Anninos:1991un,fei1992,goodman2005,Marjaneh:2016chl,AlonsoIzquierdo:2020hyp,
Adam:2019uat}. 
Applications of kink-antikink scattering~\cite{Lizunova:2020mlg} 
include the formation of abnormal nuclei~\cite{Lee:1974ma,Boguta:1983uz}, 
domain walls in crystals~\cite{PhysRevB.18.3897,Yamaletdinov:2017dlz}, folding of protein
chains~\cite{PhysRevE.83.061908, PhysRevE.83.041907}, and molecular 
dynamics~\cite{BISHOP1980955,RICE1980487}.

We are interested in examining how quantum effects modify kink scattering.
The inclusion of quantum effects presents some challenges as defects are described 
as classical objects, while other excitations of interest are quantum in nature. As a result, 
one has to couple classical and quantum degrees of freedom and to include the
quantum backreaction on the classical degrees of freedom. Fortunately much work
has been done in this area and now there is a convenient framework called the
``classical-quantum correspondence'' (CQC) in which certain quantum systems can 
be solved in terms of a classical system of 
equations~\cite{Vachaspati:2018llo,Vachaspati:2018hcu,Vachaspati:2018pps,Mukhopadhyay:2019hnb}. 
Quantum backreaction
on classical systems is in general difficult to resolve satisfactorily and is also 
mired in interpretations of quantum mechanics. However, the semiclassical 
approximation offers a path forward and we utilize it in this work.

There are only two outcomes possible in the scattering of a kink and an antikink in the \lp4 model: 
(1) the formation of a bound state (``bion'') and eventual annihilation, and (2) the kink and the antikink 
reflect and escape to infinity. However, detailed analysis~\cite{Campbell:1983xu,Anninos:1991un,AlonsoIzquierdo:2020hyp} shows that which of these outcomes
occurs depends sensitively on the initial scattering velocity, although at a coarse level, low initial velocity
kinks annihilate and high initial velocity kinks reflect. For intermediate velocities, either outcome can result
depending on the precise value of the initial velocity: there are windows of low initial velocities 
for which the kink-antikink reflect off each other instead of annihilating and there is a ``fractal''
structure in the space of initial velocities\footnote{This is an abuse of the word ``fractal'' 
since that generally involves some self-similarity. The structure here is similar to the stability
bands in solutions of the Mathieu equation~\cite{morse1953methods}.}.
The explanation of this non-trivial dependence involves the resonant energy transfer mechanism 
amongst the internal modes (the zero mode and the shape mode) of the kink and antikink. 
Recent analyses in which the classical field dynamics is truncated to just the kink translation 
and shape modes shows good agreement with the full field evolution~\cite{Manton:2021ipk,Adam:2021gat}.
Quantum effects in topologically non-trivial kink backgrounds have received significant 
attention in the past~\cite{Rajaraman:1982is,Vachaspati:2006zz} and more recently 
in~\cite{Evslin:2019xte,Evslin:2022wyx,Liu:2023dqt}. Here we are interested in quantum
effects in the zero topological charge sector as we have a background of both a kink
and an antikink.

The minimal scheme to study quantum effects during kink-antikink scattering is to include
quantum fluctuations of the $\phi$ field of the \lp4 model itself.
In other words, we only have one scalar field $\phi$ that is decomposed into a classical 
kink-antikink background ($\phi_c$) and quantum excitations on top of this background ($\hat\phi$),
\be
\phi = \phi_c + {\hat \phi}.
\label{phisplit}
\ee
A straightforward analysis of this system, however, runs into trouble -- we find that $\phi_c \to 0$
with time. The difficulty can be traced to 
the use of the semiclassical approximation in a double well potential since the wavefunctional is
not Gaussian and, in fact, may be bimodal. It is possible that some of the problems we encountered
may be due to the use of a finite lattice.
The full resolution of the difficulties is not clear to us. Hence we study quantum effects of the vacuum 
of a second scalar field $\psi$ on kink-antikink
scattering of the {\it classical} field $\phi$. The main result of our analysis is the change in
the fractal structure of the scattering as a function of the interaction strength between 
$\phi$ and $\psi$.

In Sec.~\ref{sec:formalism} we setup the model and discuss the known results for the
classical scattering of kink-antikink in the \lp4 model. We include the quantum field in
Sec.~\ref{subsec:quantdyn} and we describe the CQC. Finally in Sec.~\ref{subsec:withbkrxn} we 
solve the entire system, including backreaction. Our results are summarized in Sec.~\ref{sec:res}. We discuss the main 
conclusions along with the importance and future prospects related to this work in 
Sec.~\ref{sec:conclsn}. We work in natural units, i.e., $\hbar = c = 1$.

\section{Dynamics of the classical background}
\label{sec:formalism}
We consider a real classical scalar field $\phi$ in $1+1$ dimensions in the \lp4 theory. The Lagrangian density for the system is 
\be
\label{eq:lagden}
\mathcal{L}_\phi = \frac{1}{2} \dot{\phi}^2 - \frac{1}{2} \phi^{\prime 2} -\frac{\lambda}{4} \big(\phi^2 - \eta^2 \big)^2\,,
\ee
where, $\lambda$ and $\eta$ are parameters of the theory. The \lp4 potential has two minima corresponding to 
$\phi=\pm \eta$. The Lagrangian density in Eq.~\eqref{eq:lagden} is invariant under the transformation, $\phi \rightarrow -\phi$ and hence has a \emph{reflectional} $Z_2$ symmetry\footnote{The kinks (or antikinks) of this model are hence also known as $Z_2$ kinks (or antikinks).}. The dynamics of the field $\phi$ has been well-studied in the literature~\cite{Campbell:1983xu,Anninos:1991un,Vachaspati:2006zz} and is given by the
equation of motion,
\be
\label{eq:phieom}
\ddot{\phi} - \phi^{\prime \prime} + \lambda (\phi^2 - \eta^2)\phi = 0
\ee
By suitable rescalings, the parameters $\lambda$ and $\eta$ can be eliminated from the equation
of motion, effectively setting $\lambda=1$ and $\eta=1$. From the Lagrangian density in Eq.~\eqref{eq:lagden} the conserved energy of the classical background configuration ($E_\phi$) is given by,
\be
E_\phi = \int dx \left[ \frac{1}{2} \dot{\phi}^2 + \frac{1}{2} \phi^{\prime 2} + \frac{\lambda}{4} \big(\phi^2 - \eta^2 \big)^2 \right]\,.
\ee
The equation of motion admit kink (and antikink) solutions $\phi_{0 K (\bar{K})} (t,x)$. One can easily Lorentz boost these solutions to obtain,
\be
\phi_{0 K (\bar{K})} (t,x) = \pm \eta  \tanh{\Bigg( \eta\ \sqrt{\frac{\lambda}{2}}\  \gamma (x-v t)\Bigg)}\,,
\ee
where, the sign in front corresponds to the kink (antikink) solution 
which moves in the positive $x$ direction with a constant velocity $v$ and with Lorentz factor 
$\gamma = 1/\sqrt{1 - v^2}$. The total energy of such a configuration is given by,
\be
\label{eq:singkinken}
E_{0 K (\bar{K})} =  \frac{2 \sqrt{2}}{3} \gamma  \sqrt{\lambda} \, \eta^3 \,.
\ee

We are interested in scattering a kink and an antikink. Unlike in the integrable sine-Gordon model, the $\lambda\phi^4$ model does not 
have a closed form solution in which both a kink and an antikink are present. However, assuming that the kink-antikink start out far enough from each 
other the interaction energy between them is minimal and we can construct a field configuration of the following form,
\ba
\label{eq:phibkg}
\phi_{K\bar{K}}(t,x) &=&  \eta \left[ \tanh{\Bigg( \eta\ \sqrt{\frac{\lambda}{2}}\  \gamma (x+\Delta)\Bigg)} \right.  \nn \\
&-&  \left. \tanh{\Bigg( \eta\ \sqrt{\frac{\lambda}{2}}\  \gamma (x-\Delta)\Bigg)} + 1 \right]\,,
\ea
where, the kink and the antikink are initially displaced by a distance, 
$\Delta = v_{\rm in}t_0$, where $t_0$ is negative (and is set to 
$-100$ in our simulations)\footnote{The unit of time is
$(\sqrt{\lambda} \eta)^{-1}$, and is set to 1.}. 
The kink collision occurs at $t=0$. The above configuration has a kink and an antikink 
moving towards each other with an initial velocity $v_{\rm in}$. When the kink and the antikink 
get close and start to interact, the above configuration is no longer appropriate. 
One then needs to solve the equation of motion in Eq.~\eqref{eq:phieom}, with initial conditions specified 
by Eq.~\eqref{eq:phibkg} to get the dynamics of the field $\phi$. 
The kink-antikink pair collides, and either forms a bound state (which is also called a bion) that eventually decays into radiation or are reflected 
and travel away from each other. This outcome depends on the initial velocity $v_{\rm in}$ and 
the non-integrability of the \lp4 model makes it difficult to intuit about this dependence. Hence 
numerical simulations need to be used, the details of which are discussed in Appendix.~\ref{appsec:numerics}.
\subsection{Resonance structures in classical \lp4 kink-antikink scattering}
\label{subsubsec:class_res}
\begin{figure}
\includegraphics[width=0.49\textwidth]{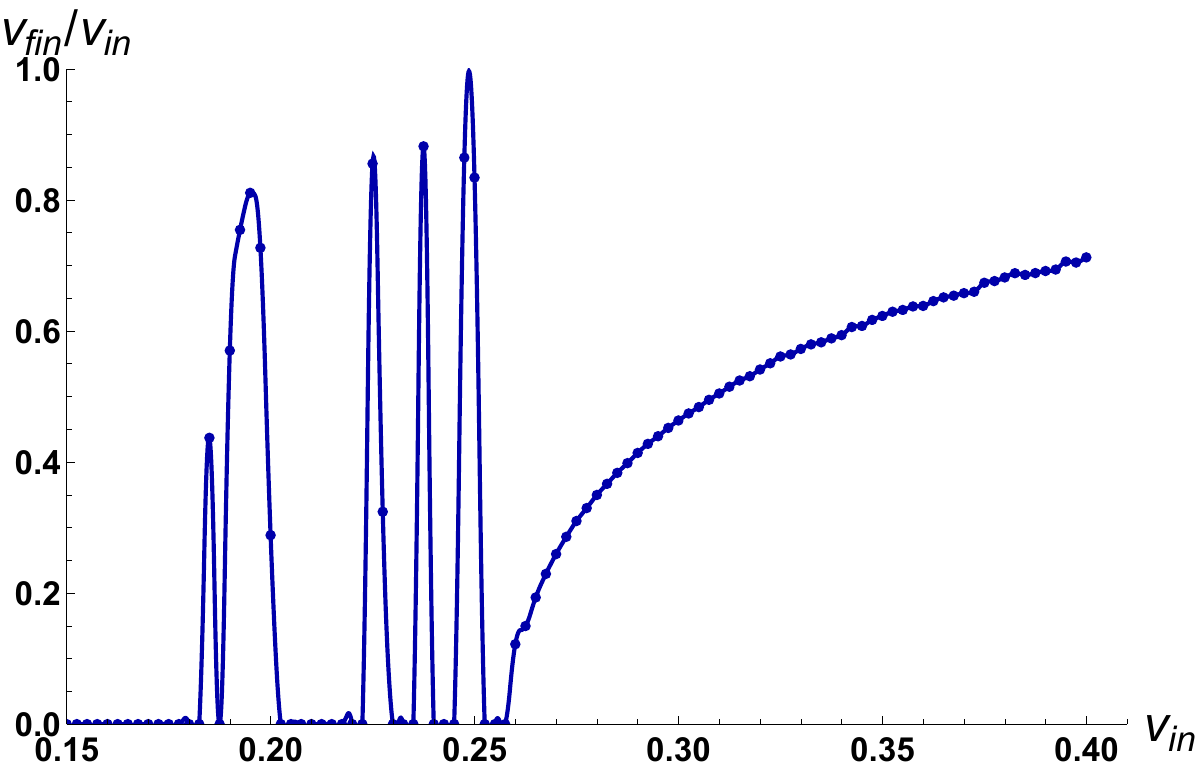}
\caption{\label{fig:resonance_class} The ratio of the final velocity to the initial velocity ($v_{\rm fin}/v_{\rm in}$) 
to the initial velocity ($v_{\rm in}$). 
Note that $v_{\rm fin} = 0$ denotes the formation of a bound state.
The curves interpolate between the actual data points. 
The parameters used are: 
$\lambda=1$, $\eta=1$, $L=100$, $N=500$, and $t_0=-100$.
}
\end{figure}
%
If a bion does not form, the kink and antikink reflect back with a definite final velocity $v_{\rm fin}$.
If a bion does form and the kink-antikink annihilate, we define $v_{\rm fin}=0$. The ratio of the final velocity 
of the kinks to the initial velocity ($v_{\rm fin}/v_{\rm in}$) depends on 
the initial velocity ($v_{\rm in}$) and
is shown in Fig.~\ref{fig:resonance_class}, a result that has been obtained and discussed extensively in
Refs.~\cite{Campbell:1983xu,AlonsoIzquierdo:2020hyp}. We present it here using our numerical calculations 
as a benchmark and most importantly to compare with the evolution when quantum backreaction is included.
Similar to prior literature, we notice the fractal structures in the plot, which indicate 
certain initial velocities for which the kink and the antikink reflect and other windows of initial
velocity in which the kinks annihilate. The annihilation windows start out being broad 
but become thinner as the initial velocity increases. 
Finally from $v_{\rm in} \sim 0.26$ the windows cease to form and we have reflection for all
$v_{\rm in} \gtrsim 0.26$.
In Ref.~\cite{AlonsoIzquierdo:2020hyp} the authors extended the initial velocities to 
$v_{\rm in} = 0.9$ but did not see any bound state formation post $v_{\rm in} \sim 0.26$.

This fractal structure has been extensively studied and well-understood in the literature. 
The kinks (or antikinks) in the \lp4 model have two distinct excitation eigenmodes.
One of them corresponds to translations of the kink, known as the zero mode, and the 
other to the internal vibrations of the kink (or antikink) known as the shape 
mode~\cite{Rajaraman:1982is, Vachaspati:2006zz}. 
In addition, there is a continuum of radiative modes on the kink background.

When the kink and the antikink collide the kinetic energy is redistributed between these modes and some of it is radiated.
In cases, where the kinetic energy is reduced to an extent where the kink-antikink do not have enough energy 
left, leads to the formation of a bound state, whereas if the decrease is not enough, they separate and reflect 
off each other. However, in some cases there is a \emph{resonant energy transfer} where the shape mode 
returns some of the kinetic energy to the zero mode leading the kink and the antikink to separate and not 
form a bound state. This accounts for the windows we see in Fig.~\ref{fig:resonance_class}. It may be worthwhile to note here given infinite precision, one would find infinite number of subsequent narrower 
windows for $v_{\rm in} \lesssim 0.26$~\cite{AlonsoIzquierdo:2020hyp}. 
\section{Dynamics of the quantum field without backreaction}
\label{subsec:quantdyn}
We now focus on the dynamics of the quantum field. It is evident that since the quantum field is coupled to 
the classical background, the non-adiabatic dynamics of the background will lead to particle production of
the quantum field. 
We will use the well-studied ``classical-quantum correspondence'' 
(CQC)~\cite{Vachaspati:2018llo,Vachaspati:2018hcu,Vachaspati:2018pps,Olle:2019skb,
Mukhopadhyay:2019hnb,Mukhopadhyay:2021wmu,Vachaspati:2018pps} technique to address 
this problem. 

The full Lagrangian density for the background $\phi$
coupled to a quantum field $\psi$ in $1+1$ dimensions is given by,
\begin{align}
\label{eq:fulllagden}
\mathcal{L} &=  \mathcal{L}_\phi
+ \frac{1}{2} (\partial_\mu \psi)^2 - \frac{1}{2}\mu^2 \psi^2 - \frac{\xi}{2} \phi^2 \psi^2\,,
\end{align}
where, 
$\mu$ is the mass of the quantum field, $\xi$ is the coupling strength, and $\mathcal{L}_\phi$ is defined in Eq.~\eqref{eq:lagden}. The truncated Lagrangian density from Eq.~\eqref{eq:fulllagden} for the quantum 
field $\psi$ can be written as,
\ba
\mathcal{L}_\psi = \frac{1}{2} \dot{\psi}^2 &-& \frac{1}{2} \psi^{\prime 2}
- \frac{1}{2} \mu^2 \psi^2 -\frac{\xi}{2}\phi^2 \psi^2\,,
\ea
where, the time-dependent background $\phi(t,x)$ is given by Eq.~\eqref{eq:phieom} if we ignore the backreaction of the quantum field on the classical background. 
Taking a closer look at the above equation it is evident that the Lagrangian density represents 
a free scalar field with a space- and time-dependent mass-squared $M^2(t,x)$,
\be
\label{eq:masssq}
M^2(t,x) = \mu^2 +  \xi\phi^2(t,x) \, .
\ee
The faster this term changes the more particle production occurs for the quantum field. Just like for the classical background, we require numerical simulations for having a quantitative understanding of the dynamics of the quantum field. We now turn to doing that.

To begin with, the spatial dimension $x$ is discretized on a 
\emph{circular} lattice of length $L$. We consider $N$ evenly spaced points on the lattice, leading to a lattice spacing of $a = L/N$. The discretized field values are defined as, $\phi_i = \phi(t,-L/2+ia)$ and $\psi_i = \psi(t,-L/2+ia)$, where, $i=1,2,\dots,N$. The discretized Lagrangian will now be,
\ba
L_{\psi, \mathrm{disc.}} &=& a \sum_{i=1}^N \left[ \frac{1}{2} \right. \dot{\psi}_i^2 
+ \frac{1}{2} \psi_i \Bigg( \frac{\psi_{i+1}-2\psi_i+\psi_{i-1}}{a^2} \Bigg)  \nn \\
&-& \frac{1}{2} \mu^2 \psi_i^2 
-  \left. \frac{\xi}{2} \phi_i^2 \psi_i^2\right]\,.
\ea
Note that we have performed a spatial integration by parts in writing the gradient term above. Let us now switch to a more compact matrix notation, where we can define a column vector for the discretized $\psi$ field as, $\bm \psi = (\psi_1, \psi_2, \dots, \psi_N)^T$. In this notation, the discretized Lagrangian ($L_{\bm \psi, \mathrm{disc.}}$) is,
\be
\label{eq:discL}
L_{\bm \psi, \mathrm{disc.}} = \frac{a}{2} \dot{\bm \psi}^T \dot{\bm \psi} - \frac{a}{2} \bm \psi^T \bm \Omega^2 \bm \psi\,,
\ee
where $\bm \Omega^2$ is an $N \times N$ matrix defined as,
\be
\label{eq:omegasq}
[\bm{\Omega}^2]_{ij} = 
\begin{cases}
+{2}/{a^2}+M^2(t,x),& i=j\\
-{1}/{a^2},& i=j\pm1\ (\text{mod}\ N)\\
0,&\text{otherwise}\,,
\end{cases}
\ee
where, $M^2(t,x)$ is defined in Eq.~\eqref{eq:masssq} and we now replace $\phi$ by the discretized value of 
$\phi$ at a point $i$, $\phi_i$. The conjugate momentum $\bm \pi$ can be calculated from Eq.~\eqref{eq:discL} which allows us to define the discretized Hamiltonian ($H_{\bm \psi, \mathrm{disc.}}$) as,
\be
\label{eq:dischamil}
H_{\bm \psi, \mathrm{disc.}} = \frac{a}{2}\bm \pi^T \bm \pi + \frac{a}{2} \bm \psi^T \bm \Omega^2 \bm \psi\,.
\ee
We now move on to quantizing the theory. This is achieved in the Heisenberg picture by promoting the discretized field value $\psi_i$ to an operator $\hat{\psi}_i$ at the lattice point $i$. Following~\cite{Vachaspati:2018llo,Vachaspati:2018hcu}, we introduce the complex time-dependent matrix $\bm Z$. The elements of $\bm Z$, $Z_{ij}$ satisfy the following relation,
\be
\hat{\psi}_i = Z_{ij}^* \hat{a}_{j} (t_0) + Z_{ij} \hat{a}_{j}^{\dagger}(t_0)\,,
\ee
where, we assume that the background is static for any time $t \leq t_0$.
The complete dynamics of $\hat{\psi}_i$ is given by,
\be
\label{eq:zeqn}
\ddot{\bm Z} + \bm \Omega^2 \bm Z = 0\,,
\ee
with the specific initial conditions,
\be
\label{eq:zinit}
\bm{Z}(t_0) = -\frac{i}{\sqrt{2 a}} \bm{\Omega}(t_0)^{-1/2}, \ 
\dot{\bm{Z}}(t_0) = \frac{1}{\sqrt{2 a}} \bm{\Omega}(t_0)^{1/2}.
\ee
The above initial conditions come with a small caveat. The initial vacuum for the quantum field is chosen 
around a static background that is given by the initial conditions, but the background is not static and has some time dependence due to the initial velocities of the kinks.
This leads to some unwanted but small excitations in the $\psi$ field, which are radiated out even before
the kinks have collided. This gives a small error $\lesssim 0.1\%$ of the total energy in the initial conditions for the range of initial velocities $v_{\rm in}$ we have investigated.
\section{Dynamics with backreaction}
\label{subsec:withbkrxn}
We now focus on the main aspect of this work - the dynamics of the classical background coupled to the quantum field including backreaction. In this scenario, the dynamics of the classical background $\phi$ also gets a contribution from the $\psi$ field,
\be
\ddot{\phi}-\phi^{\prime \prime} + [ \lambda (\phi^2 - \eta^2) + \xi \psi^2 ] \phi= 0\,.
\ee
To take the effects of the quantum field $\psi$ into account we can use the semiclassical 
approximation in which $\psi^2$ is replaced by 
$\langle 0 |\hat{\psi}^2 |0 \rangle \equiv \langle \hat{\psi}^2 \rangle$, where 
the expectation value is calculated in the vacuum state of the quantum field $\hat \psi$.
(States do not evolve in the Heisenberg picture.) 
The background equation can now be written as,
\be
\ddot{\phi}-\phi^{\prime \prime} + 
[ - m_b^2  + \lambda \phi^2 + \xi \langle \hat{\psi}^2  \rangle ] \phi= 0\,.
\label{sceqphi}
\ee
where $m_b^2 \equiv \lambda \eta^2$. The parameters in \eqref{sceqphi} should
be thought of as bare parameters that will get renormalized due to the quantum
field $\hat\psi$. We only need to consider ``mass renormalization'' to the
order we are working in and we can write,
\be
\ddot{\phi}-\phi^{\prime \prime} + 
[ - m^2  + \lambda \phi^2 
+ \xi (\langle \hat{\psi}^2\rangle - \langle \hat{\psi}^2 \rangle_{\phi = \eta}  ) ] \phi= 0\,.
\label{sceqphiren}
\ee
where the subscript $\phi=\eta$ implies that the expectation value is to be taken
in the ground state of $\hat\psi$ in the trivial background $\phi=\eta$.
Now $m^2$ is the physical mass parameter and is related to $m_b^2$ by
\be
m^2 = m_b^2 - \xi \langle \hat{\psi}^2 \rangle_{\phi = \eta}.
\label{renmass}
\ee

In its discretized form Eq.~\eqref{sceqphiren} is,
\ba
\label{eq:bkrxnphi}
&&
\ddot{\phi}_i - \frac{1}{a^2}( \phi_{i+1} - 2 \phi_i + \phi_{i-1} ) 
\nn \\
&& \hskip -0.5 cm
+\biggl [ -m^2 + \lambda \phi_i^2 
+ \xi \sum_{j=1}^N \big( |Z_{ij}|^2 -  |Z_{ij}|^2_{\phi=\eta} \big) \biggr ] \phi_i = 0.
\label{eq:bkrxnphi}
\ea
where we have used~\cite{Vachaspati:2018hcu},
\be
\langle  \hat{\psi}_i^2 \rangle = \sum_{j=1}^N |Z_{ij}|^2\, .
\ee
Thus, Eq.~\eqref{eq:bkrxnphi} has to be solved for $\phi_i$ with initial conditions,
\ba
\label{eq:phiinit}
\phi_i(t_0) &=& \phi_{K\bar{K}}(t_0,-L/2+ia)\,,\nn \\
\dot{\phi}_i(t_0) &=& \dot{\phi}_{K\bar{K}}(t_0,-L/2+ia)\,,
\ea
where, as before, $i=1,2,\dots,N$ and $\phi_{K\bar{K}} (t,x)$ is given by Eq.~\eqref{eq:phibkg}. 

The complete backreacted dynamics is given by Eqs.~\eqref{eq:bkrxnphi} and~\eqref{eq:zeqn}, 
with initial conditions given by Eqs.~\eqref{eq:phiinit} and~\eqref{eq:zinit}.
\subsection{Energy}
\label{subsec:energy}
The conserved total energy of the system ($E$) is given by,
\be
E = E_\phi + E_\psi^{(R)}\,,
\ee
where, $E_\phi$ is the energy in the classical background and $E_\psi^{(R)}$ is the renormalized energy in the quantum field. The total energy in the classical background can be defined as,
\ba
\label{eq:phiendisc}
E_\phi &=& a \sum_{i=1}^N \rho_{\phi,i} = a \sum_{i=1}^N \left[ \frac{1}{2}\dot{\phi}_i^2 +\frac{1}{4 a^2} \big( (\phi_{i+1}-\phi_i)^2 \big) \right. \nn \\
&& \hskip 1 cm
+(\phi_i-\phi_{i-1})^2 + \left.  \frac{\lambda}{4} \big(\phi_i^2 - \eta^2 \big)^2 \right]\,.
\ea
The energy in the quantum field is,
\be
E_\psi \equiv \langle 0|\hat{H}_{\psi, \text{disc.}} |0\rangle = \frac{a}{2} \text{Tr}\Bigg( \dot{\bm Z}^\dagger \dot{\bm Z} + \bm Z^\dagger \Omega^2 \bm Z \Bigg)\,.
\ee
The discretized energy density in $\hat \psi$ can be defined as,
\ba
&&
\rho_{\psi, i} = \sum_{j=1}^{N} \biggl [ \frac{1}{2} |\dot{Z}_{ij}|^2 
+\frac{1}{4 a^2} \biggl \{ |Z_{i+1,j}-Z_{ij}|^2 \nn \\
&& \hskip 1 cm
+|Z_{ij} -  Z_{i-1,j}|^2 \biggr \} + \frac{1}{2} \biggl \{ \mu^2 + \xi \phi_i^2 \biggr \} |Z_{ij}|^2 \biggr ].
\label{rhopsi}
\ea
Similar to the renormalized mass defined in \eqref{renmass}, the renormalized 
energy density in the quantum field may be defined as,
\be
\label{eq:endendiscpsi}
\rho_{\psi, i}^{(R)} = \rho_{\psi, i} - \rho_{\psi, i}|_{\phi=\eta}\,,
\ee
where, we subtract the energy density of the quantum field in the trivial vacuum ($\phi = \eta$) from the energy density. Hence, the renormalized total energy in the quantum field is,
\be
\label{eq:totalendendiscpsi}
E_\psi^{(R)} = E_\psi - E_\psi|_{\phi=\eta}.
\ee
\begin{figure}
\includegraphics[width=0.48\textwidth]{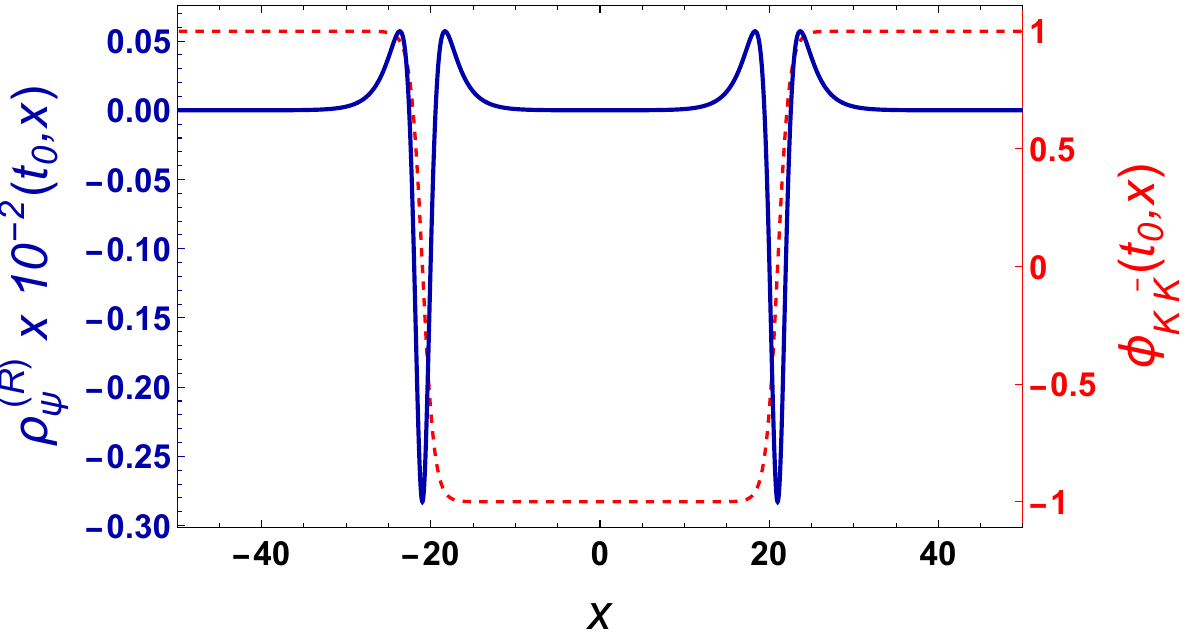}
\caption{\label{fig:ics} The initial renormalized energy density in $\psi$ (solid dark blue curve)  and the initial profile of the kink-antikink background $\phi$ (dashed red curve). We note that the kink and the antikink is dressed with the $\psi$ particles. The main parameters are: $v_{\rm in} = 0.21$, $\xi = 0.05$, $\mu = 0.1$, $\lambda=1$, $\eta = 1$, $L=100$, $N=500$, and $t_0=-100$.
}
\end{figure}
\subsection{Initial structure of the vacuum}
The initial vacuum structure of the quantum field can be visualized from Fig.~\ref{fig:ics}. 
In the figure, we show the initial renormalized energy density of $\psi$, ($\rho_{\psi, i}^{(R)} (t=t_0,x)$). 
The figure also shows the background $\phi$ at the initial 
time $\phi_{K\bar{K}}(t=t_0,x)$ (see Eq.~\eqref{eq:phiinit}). 
In the trivial vacuum, $\phi = \eta$, the renormalized energy density in $\psi$ vanishes,
as in Fig.~\ref{fig:ics} at the boundaries of the lattice.
However, at the position of the kink and 
the antikink, there is a big dip in the energy density of $\psi$ that depends on the
parameters of the model
The dips are the ground state of $\psi$ in this particular (kink-antikink) background.
As the background changes, that is, as the kink-antikink move, the dips move along with them. 
This will be evident in Sec.~\ref{subsec:results1} where we study dynamics of the background.
\section{Results}
\label{sec:res}
Our numerical methods and checks are described in Appendix~\ref{appsec:numerics} and \ref{appsec:quality}.
\subsection{Quantum kink-antikink scattering with backreaction}
\label{subsec:results1}
\begin{figure*}
\centering
\subfloat[(a)] {\includegraphics[width=0.49\textwidth]{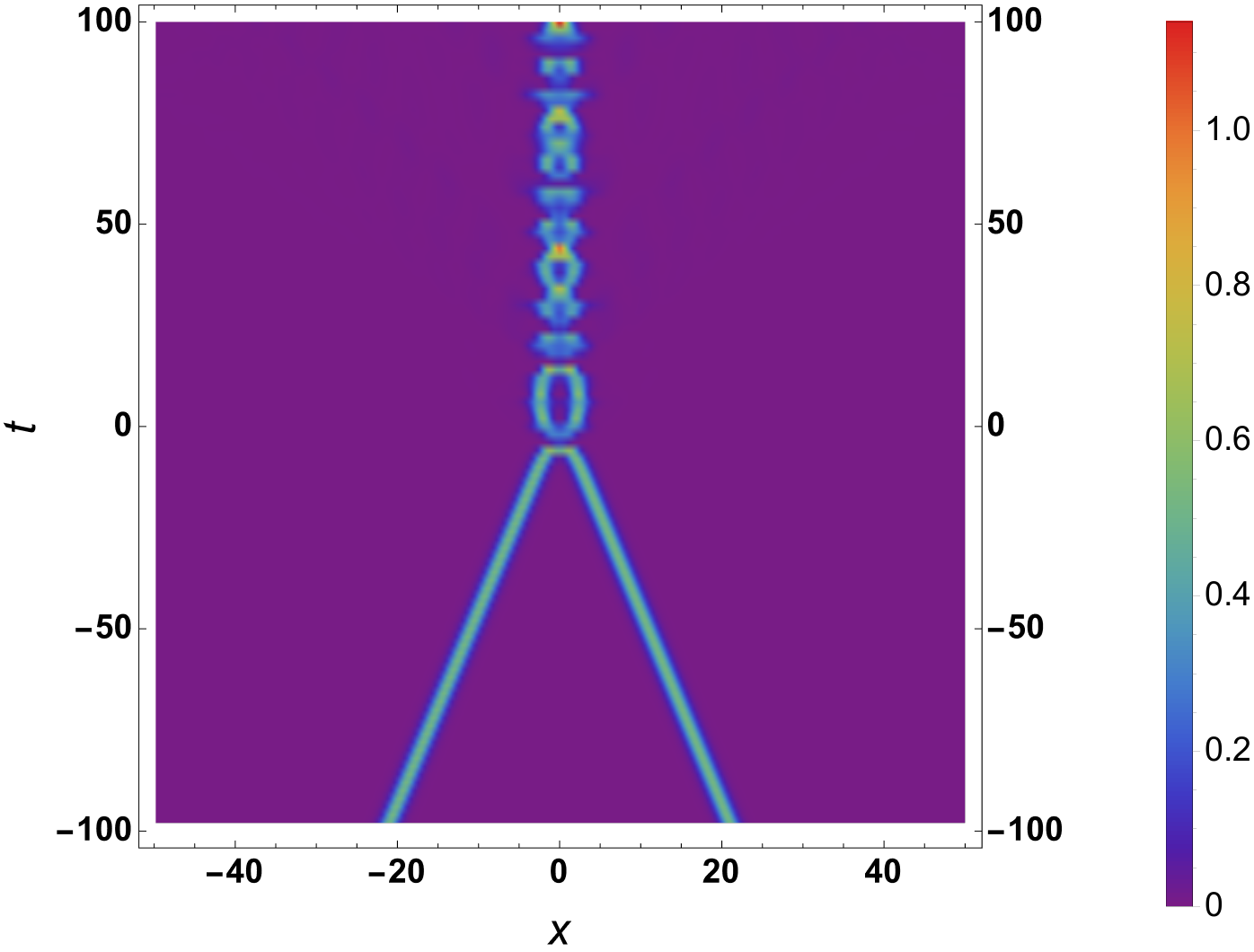}\label{fig:classphibound}}\hfill
\subfloat[(b)] {\includegraphics[width=0.49\textwidth]{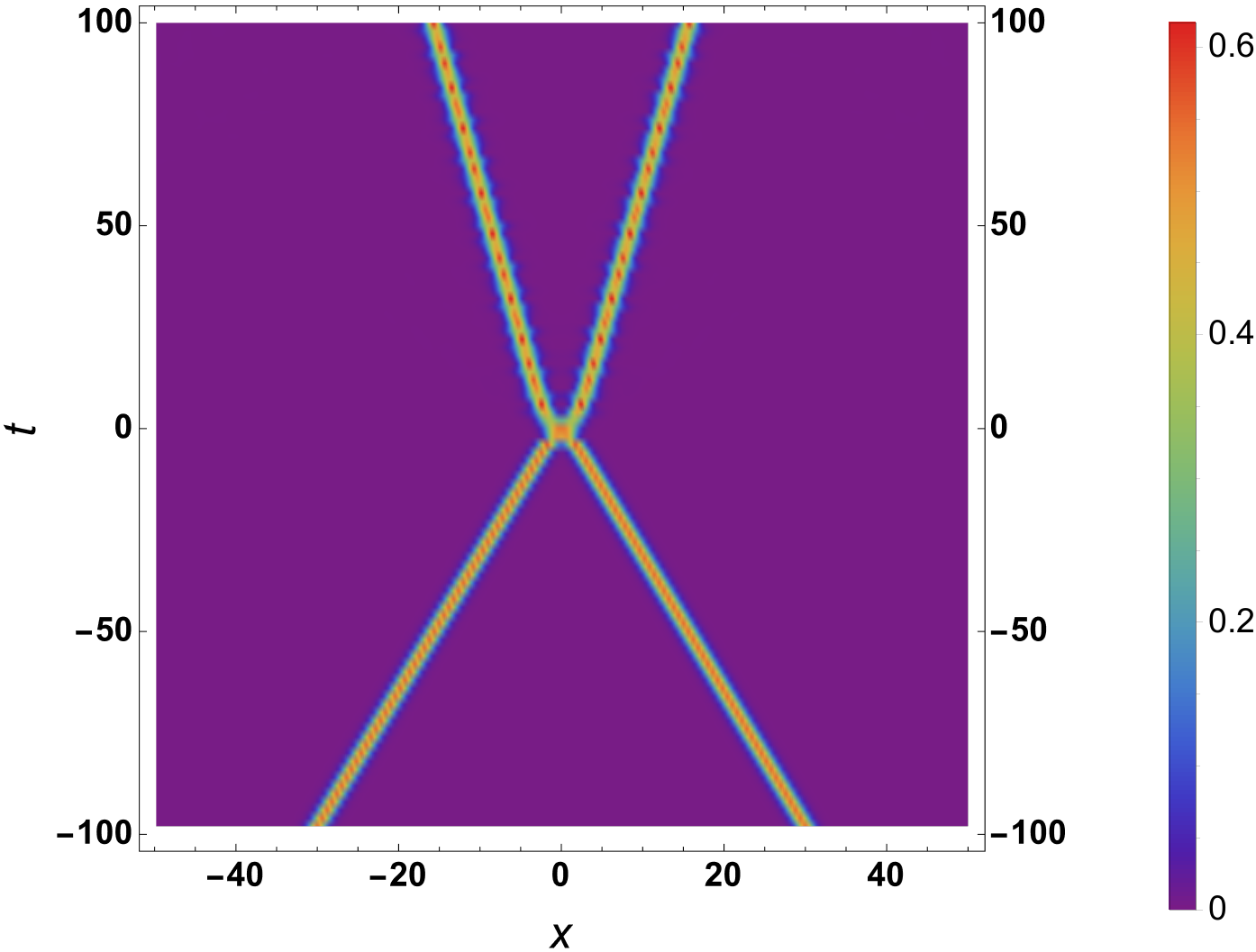}\label{fig:classphiunbound}} \hfill
\\
\subfloat[(c)] {\includegraphics[width=0.49\textwidth]{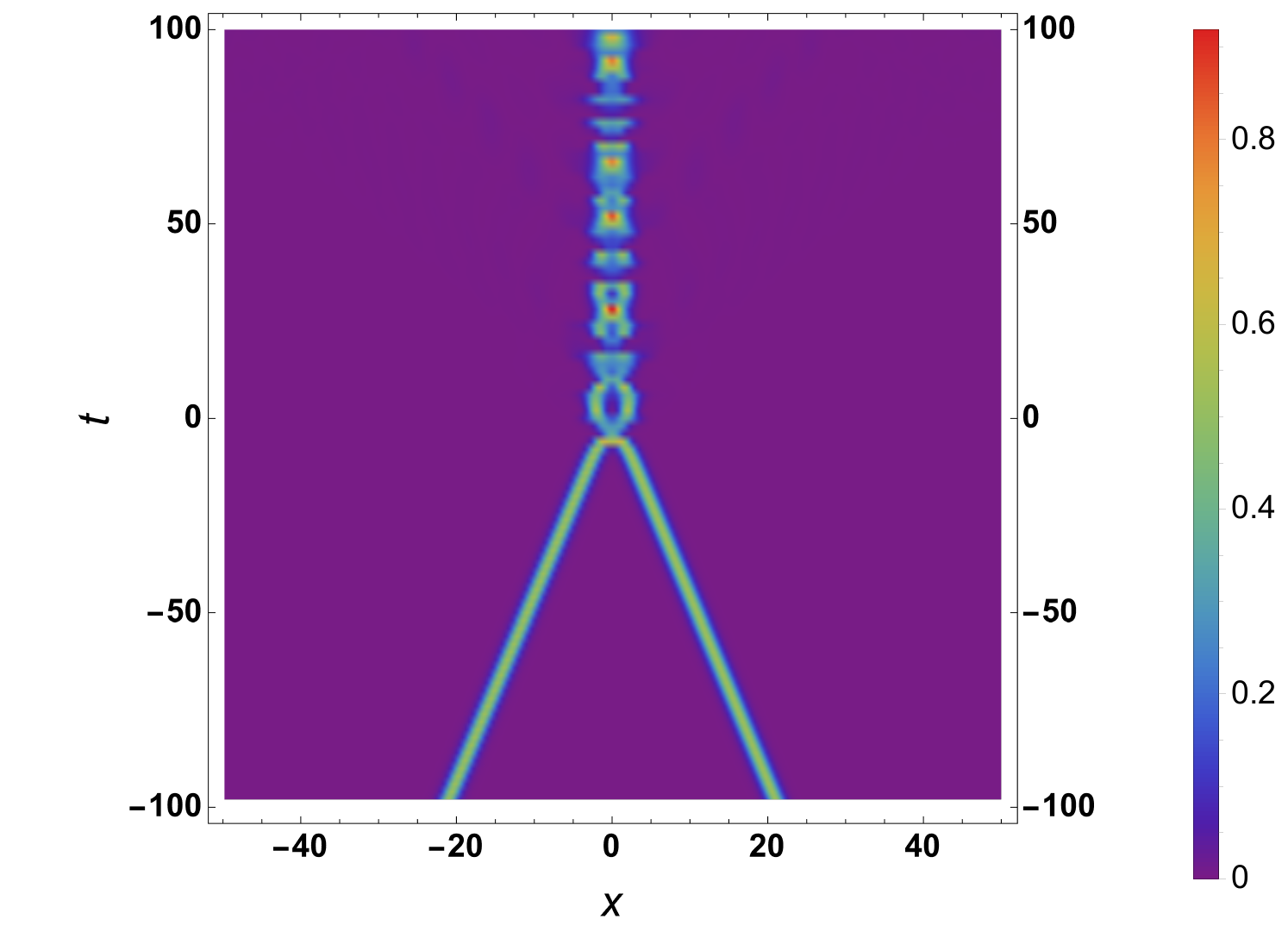}\label{fig:quantphibound}}\hfill
\subfloat[(d)] {\includegraphics[width=0.49\textwidth]{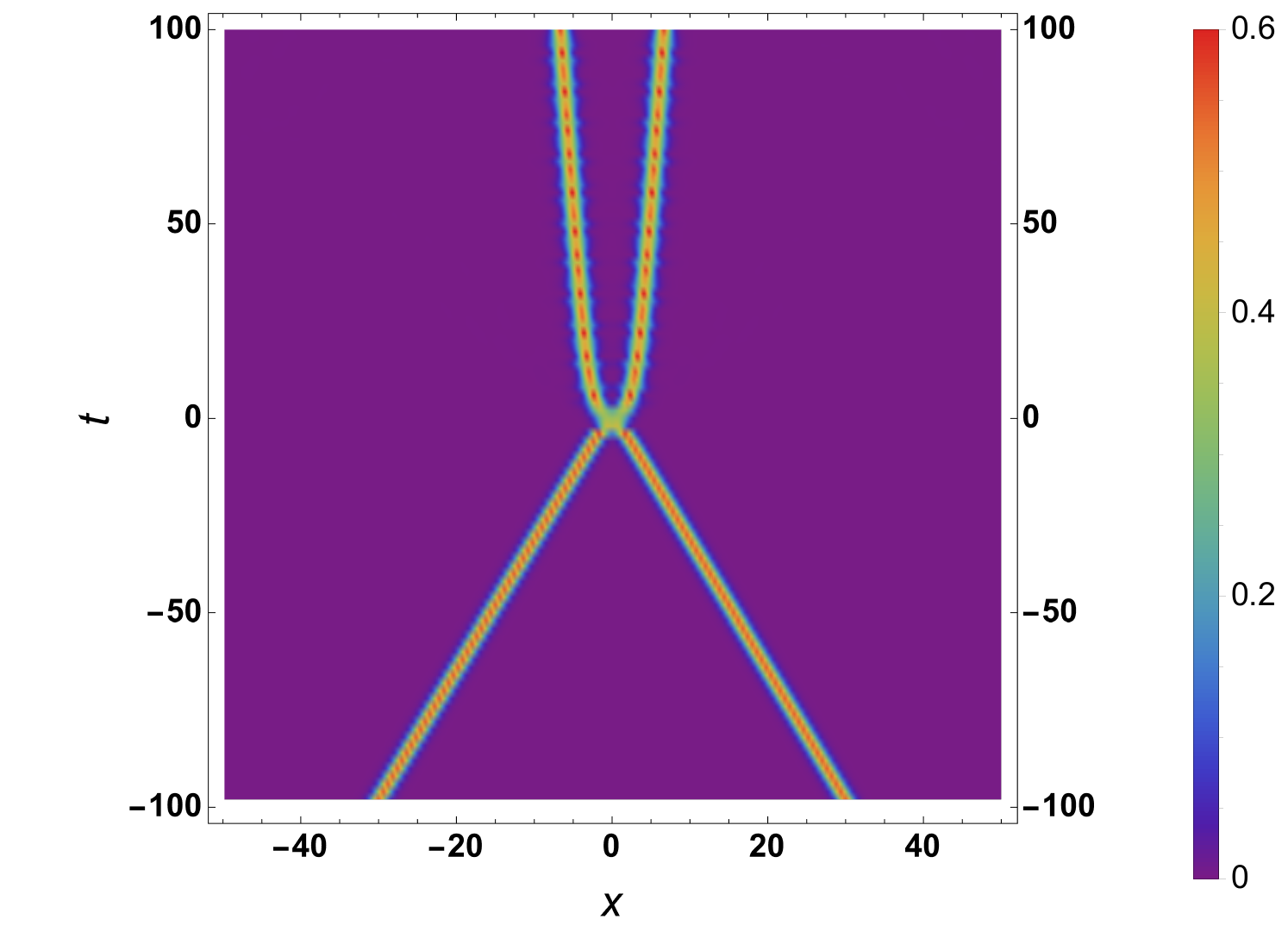}\label{fig:quantphiunbound}} \hfill
\caption{\label{fig:phi_enden} Density plots showing the time evolution of the energy density  ($\rho_{\phi,i}$) in the 
background (see Eq.~\eqref{eq:phiendisc}) - \emph{Top Panel: }Classical evolution - (a) $v_{\rm in} = 0.21$ 
where the final state is a bound state, (b) $v_{\rm in} = 0.3$ to show a case where the kink and the 
antikink get reflected post-collision. \emph{Bottom Panel: } Evolution including coupling to the quantum 
field and backreaction - (c) same as (a), (d) same as (b). The parameters are: 
$\xi = 0.05$, $\mu = 0.1$, $\lambda=1$, $\eta = 1$, $L=100$, $N=500$, and $t_0=-100$ for all the cases.
The animations corresponding to the different cases can be found at \href{https://sites.google.com/asu.edu/mainakm/animations\#h.oaaca9ogoc42}{https://sites.google.com/asu.edu/mainakm}.
}
\end{figure*}
In Fig.~\ref{fig:phi_enden} we show the energy density in the background field $\phi$ -- the expression in 
rectangular brackets in Eq.~\eqref{eq:phiendisc} -- as a function of time (vertical axis). The top panels compare 
the classical evolution ($\xi=0$) for $v_{\rm in}=0.21$, when a bion forms, and for $v_{\rm in}=0.30$, when the
kinks reflect. The bottom two panels show the corresponding evolution for ($\xi = 0.05$). The bion
is now tighter; the reflected kinks have smaller velocity. The behavior is expected since the kinks in
the $\xi \ne 0$ case excite the quantum vacuum of $\hat \psi$, produce particles, and lose energy
during the evolution. The energy in $\hat\psi$ must come from the kinetic energy of the kinks which
results in a tighter bion for $v_{\rm in}=0.21$ and for slower reflected kinks for $v_{\rm in}=0.30$.
\begin{figure}
\subfloat[(a)] {\includegraphics[width=0.49\textwidth]{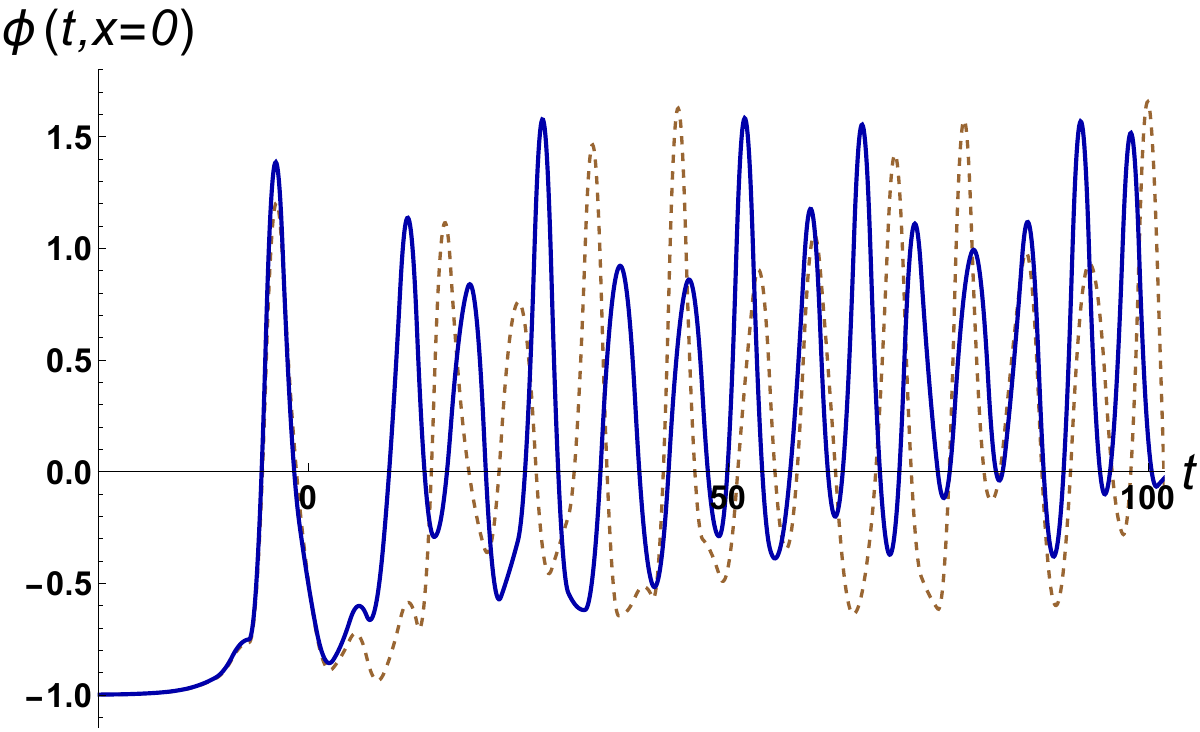}\label{fig:phi0bound}}\hfill 
\\
\subfloat[(b)] {\includegraphics[width=0.49\textwidth]{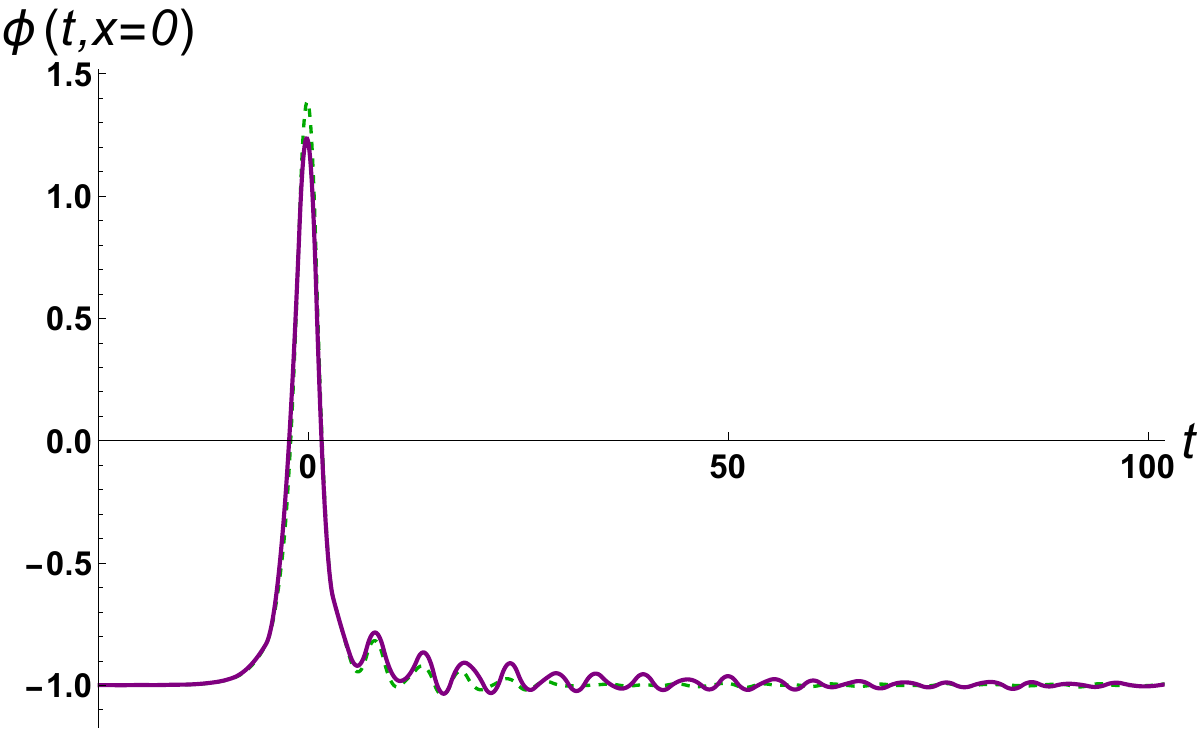}\label{fig:phi0unbound}} \hfill
\caption{\label{fig:phi0} Time evolution of the value of the background $\phi$ at the center of the lattice $\phi(t,x=0)$ contrasting the classical evolution (dashed) with the case including quantum backreaction (solid) for - (a) $v_{\rm in} = 0.21$ (brown and dark blue respectively) and (b) $v_{\rm in} = 0.3$ (dark green and purple respectively). The parameters used are the same as the ones stated in Fig.~\ref{fig:phi_enden}.
}
\end{figure}
In the case $v_{\rm in}=0.21$, where the final state is a bound state, we see that the kink and the antikink 
travel towards each other, collide and get stuck leading to a caterpillar-like structure in Fig.~\ref{fig:phi_enden}. 
Each segment of the caterpillar shows that the kinks separate for a short duration but then come back and 
collide again. This repeats for some time and finally the kink-antikink do not have enough energy to separate 
and hence lose their individual identity to form a bound state (or bion) that decays into radiation. 
It is difficult to see the difference between 
Figs.~\ref{fig:classphibound} and~\ref{fig:quantphibound}, since in both cases the caterpillar structure 
exists and repeated collisions makes it difficult to notice any differences. To clarify the differences,
we plot the value of the background $\phi$ at the center of the lattice, $\phi (t,x=0)$,
in Fig.~\ref{fig:phi0} for $v_{\rm in}=0.21$ and for $v_{\rm in}=0.30$.
\begin{figure}
\includegraphics[width=0.49\textwidth]{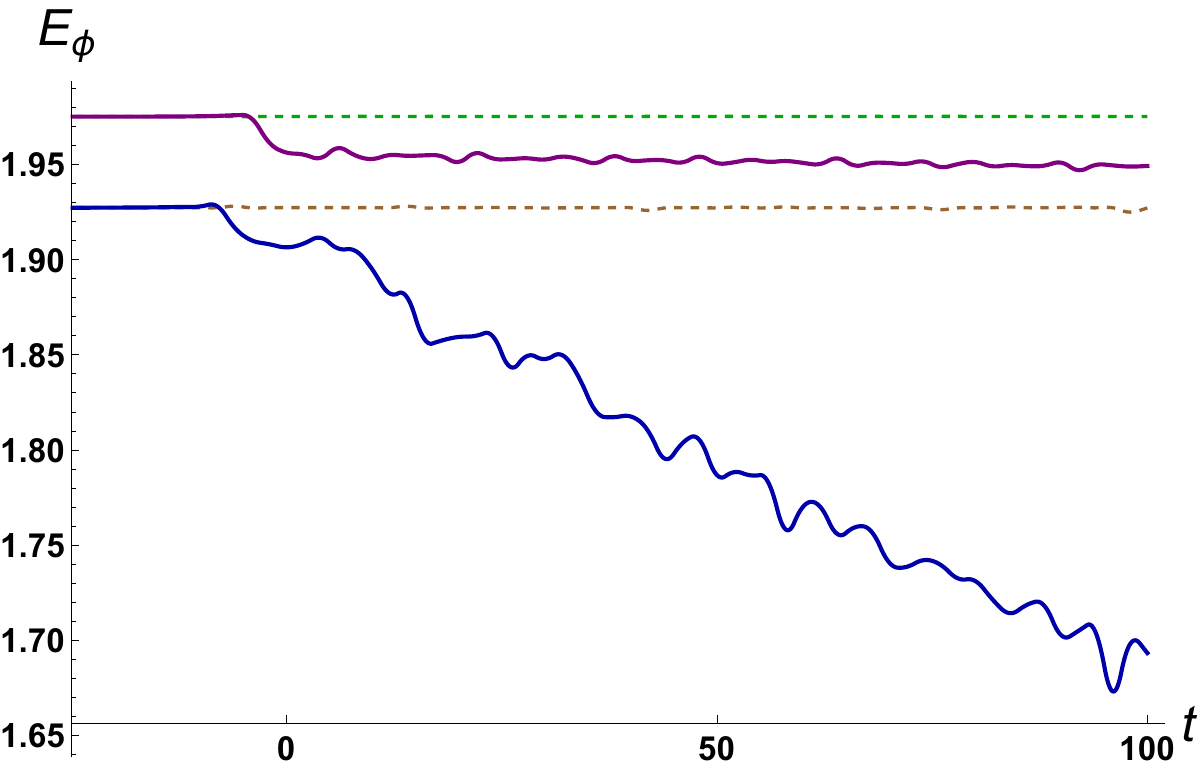}
\caption{\label{fig:enphi} Time evolution of the total energy in the background $\phi$ ($E_\phi$) for the classical evolution (dashed) and the case including quantum backreaction (solid) for - (a) $v_{\rm in} = 0.21$ (brown and dark blue respectively) and (b) $v_{\rm in} = 0.3$ (dark green and purple respectively). The parameters used are the same as the ones stated in Fig.~\ref{fig:phi_enden}.
}
\end{figure}
\begin{figure*}[ht!]
\subfloat[(a)] {\includegraphics[width=0.49\textwidth]{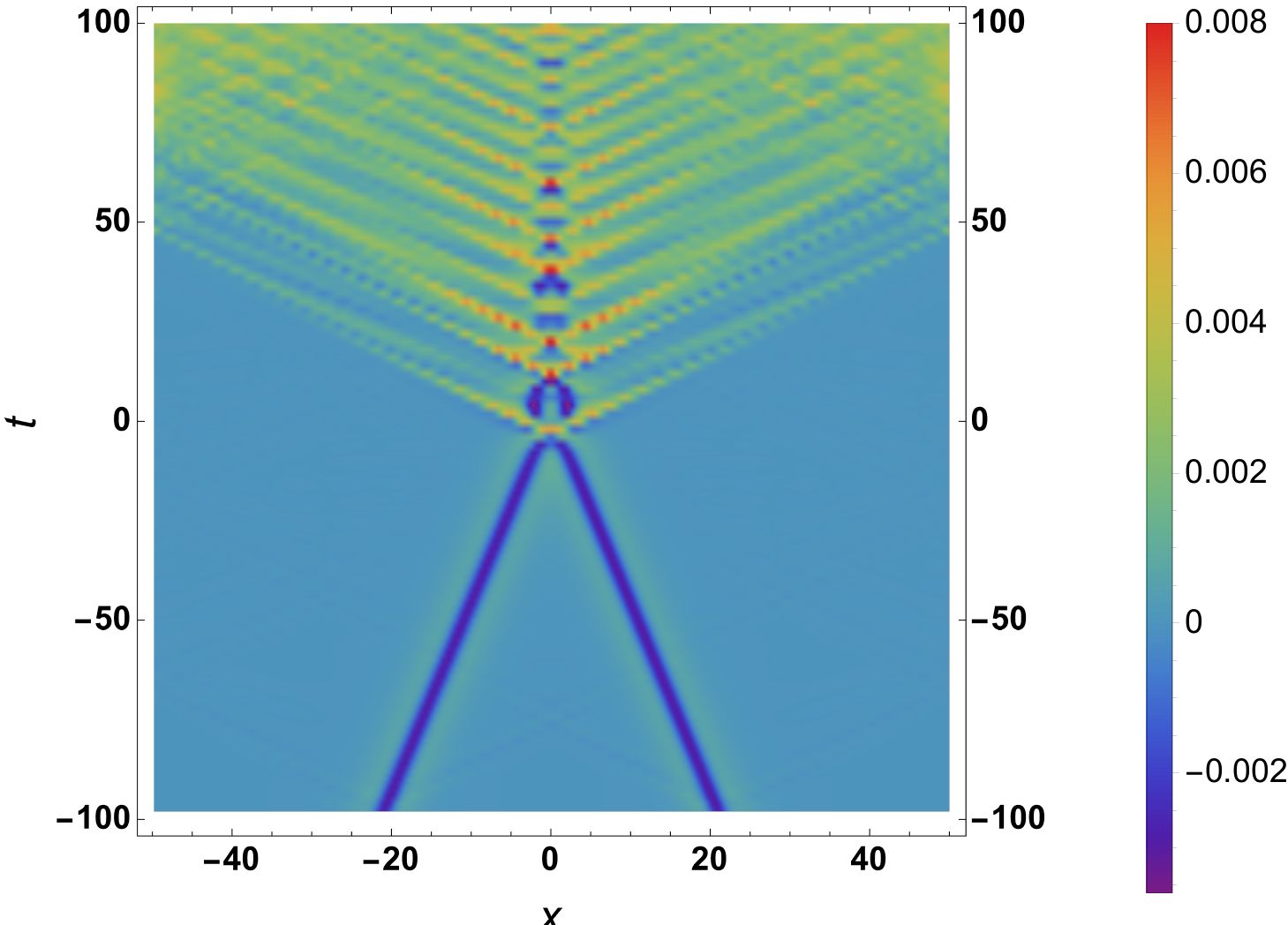}
\label{fig:quantpsibounda}}\hfill
\subfloat[(b)] {\includegraphics[width=0.49\textwidth]{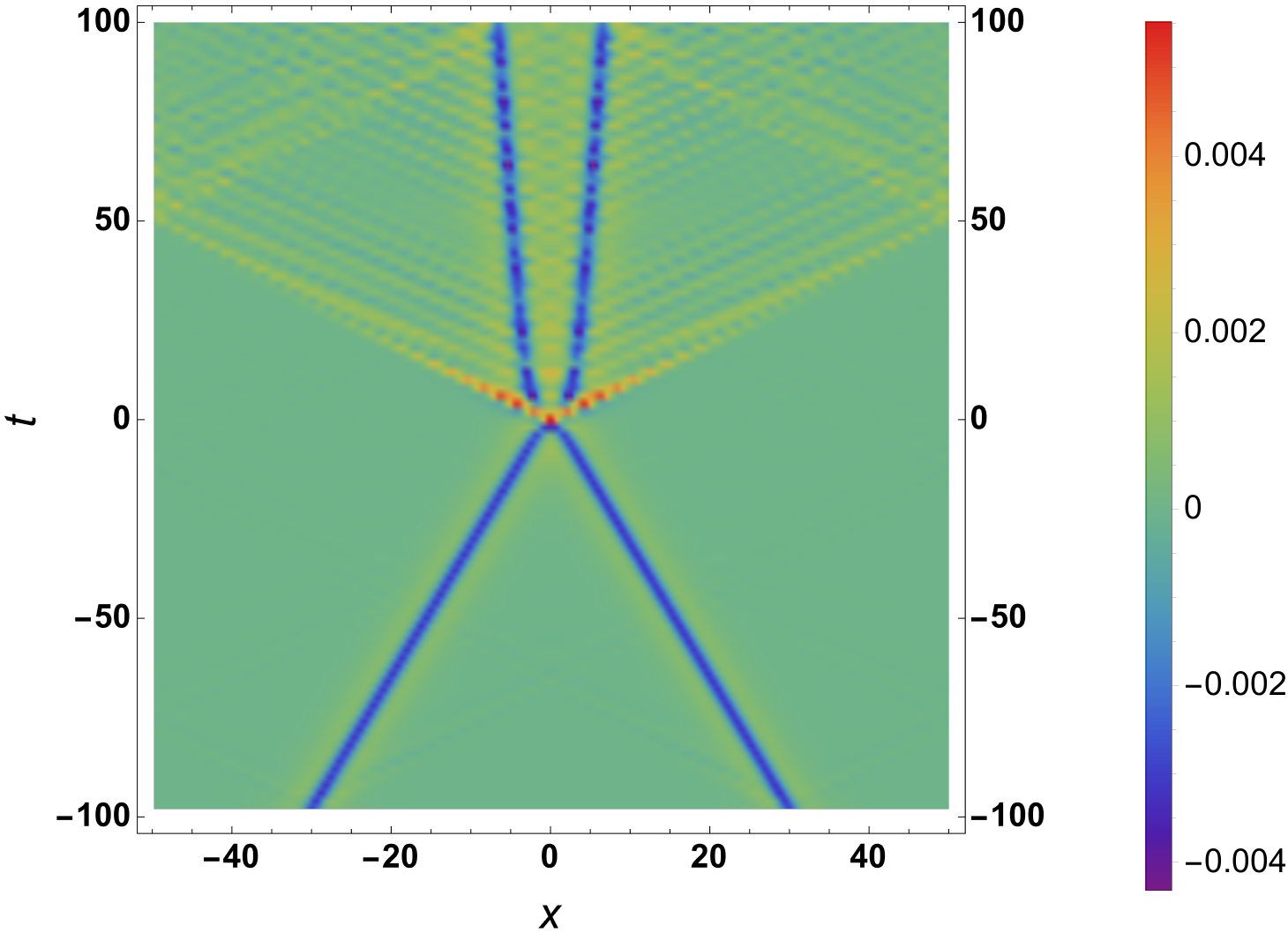}
\label{fig:quantpsiunboundb}} \hfill
\caption{\label{fig:psi_endenab} Time evolution of the renormalized energy density
$\rho_\psi^{(R)}$. (a) $v_{\rm in} = 0.21$ where a bound state is formed and 
(b) $v_{\rm in} = 0.3$ where the kink and the antikink undergo reflection. The 
parameters are the same as for Fig.~\ref{fig:phi_enden}.
The animations corresponding to the different cases can be found at \href{https://sites.google.com/asu.edu/mainakm/animations\#h.oaaca9ogoc42}{https://sites.google.com/asu.edu/mainakm}.
}
\end{figure*}
In Fig.~\ref{fig:enphi} we show the time dependence of the energy in the $\phi$ background  
(see Eq.~\eqref{eq:phiendisc}). Before the collision, the energy in the background is constant and 
is given by twice that of the single kink (or antikink) energy (see Eq.~\ref{eq:singkinken}), which is basically the sum of the individual energies 
of the kink and the antikink. For the classical evolution (dashed curves), $E_\phi$ is the total energy
and is conserved.
In the presence of the quantum field, the outcome depends on $v_{\rm in}$.
When a bound state is formed (solid dark blue curve), the repeated collisions lead to a 
cascading drop in $E_\phi$ and the lost energy goes into radiating $\psi$ particles.

The loss of energy from the kinks to $\psi$ quanta can be visualized in Fig.~\ref{fig:psi_endenab}.
The plot shows the time evolution of the renormalized energy density in $\psi$, 
denoted by $\rho_\psi^{(R)}$ (see Eqs.~\eqref{rhopsi}, \eqref{eq:endendiscpsi}).
As we have noted in Fig.~\ref{fig:ics}, there is a ``cloud'' of $\psi$ around the position of the kink. 
This cloud perfectly follows the kink and the antikink at early times as can be seen by comparing
Figs.~\ref{fig:phi_enden} and \ref{fig:psi_endenab}.

The collisions are visible as the segments of the caterpillar after $t=0$. In the $v_{\rm in}=0.21$
case, as the kinks collide, separate
and collide again, we see bursts of radiation that are produced, which then propagate outwards 
away from the center.
By virtue of periodic boundary conditions, once the radiation reaches the end of the lattice they wrap 
around and return back towards the center. 
The bursts of radiation in the case of a bound state formation is what is observed in Fig.~\ref{fig:enphi} 
(solid dark blue curve), as a cascading decrease in the total energy in the background.

When the kink-antikink pair is reflected, as in Fig.~\ref{fig:quantpsiunboundb}, there is a burst of $\psi$ radiation
at the time of collision seen as the two yellow bands moving out from the collision point in the figure
After the collision, the $\psi$ clouds keep moving with the kink and the antikink but the kink
internal modes are excited and there are some weaker bursts of $\psi$ radiation. There are
the light yellow bands being radiated out as the kinks move away from each other in 
Fig.~\ref{fig:quantpsiunboundb}. The energy loss appears as the undulating features in
the $v_{\rm in}=0.30$ curve at late times in Fig.~\ref{fig:enphi}.
\begin{figure}
\includegraphics[width=0.49\textwidth]{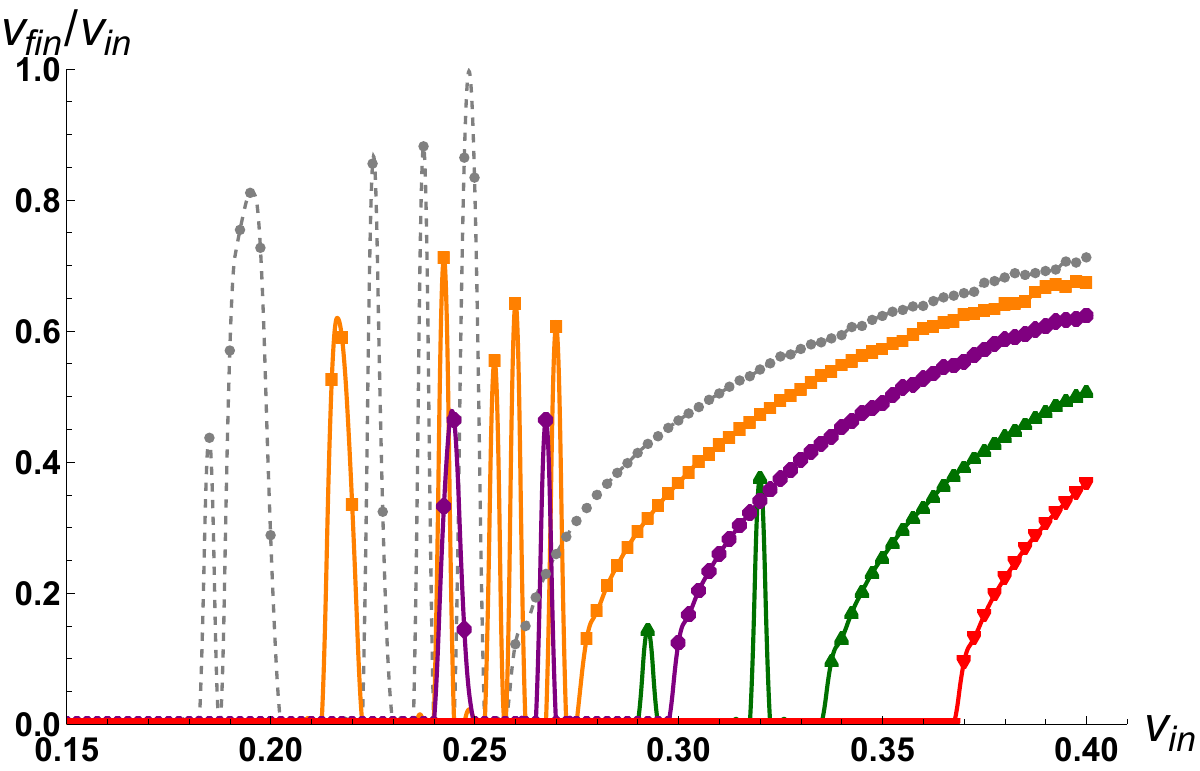}
\caption{\label{fig:resonance} The ratio of the final velocity to the initial velocity ($v_{\rm fin}/v_{\rm in}$) post-collision to the initial velocity ($v_{\rm in}$). Note that $v_{\rm fin} = 0$ denotes the formation of a bound state. The dashed gray line shows the classical evolution ($\xi = 0$), whereas the solid lines denote the evolution including quantum backreaction for different values of $\xi$: $\xi = 0.03$ (orange), $\xi = 0.05$ (purple), $\xi = 0.08$ (dark green), and $\xi = 0.1$ (red). The actual data points are also shown. The other parameters used are: $\mu = 0.1$, $\lambda=1$, $\eta=1$, $L=100$, $N=500$, and $t_0=-100$.
}
\end{figure}
Finally we come to our main objective of examining how the fractal nature of the classical scattering shown
in Fig.~\ref{fig:resonance_class} changes when the kinks scatter in the quantum vacuum of a second field.
The results are presented in Fig.~\ref{fig:resonance}. The dashed gray curve shows the results of the 
classical scattering and is the same as presented in Fig.~\ref{fig:resonance_class}. 
The other curves show the results of the scattering as we turn up the interaction strength $\xi$.
An increase in $\xi$ shifts the scattering peaks to higher $v_{\rm in}$ but also reduces the peak 
$v_{\rm fin}/v_{\rm in}$. The critical $v_{\rm in}$ above which there is no further fractal structure
also increases with increasing $\xi$. For example, for $\xi=0$ the critical $v_{\rm in}$ is $\sim 0.26$
while for $\xi = 0.1$ it is $\sim 0.37$. Some of these features can be qualitatively understood based
on our earlier discussion that the time-dependent background of the scattering kinks excites
$\psi$ radiation and causes the kinks to lose kinetic energy. This process makes it easier for
the kink-antikink to form a bound state and annihilate and it takes higher initial velocities for
the kinks to reflect. Also, in cases that the kinks do reflect, the lost energy means that the
final velocities of the kinks will be lower.

We have checked that our results are insensitive to the choice of lattice spacing, $a$,
and to the size of the simulation box, $L$. A comparison of our results for different choices 
of $a$ and $L$ is shown in Appendix.~\ref{appsec:quality}.
\section{Conclusions and Discussion}
\label{sec:conclsn}
%
While kink-antikink scattering has been thoroughly investigated in the literature~\cite{Campbell:1983xu,Anninos:1991un,AlonsoIzquierdo:2020hyp}, our focus
in this work has been on {\it quantum} effects on the scattering. Even if the kink-antikink propagate
in a quantum vacuum, they provide a time-dependent background that can produce particles and
radiation, which in turn depletes the kinetic energy of the kinks and modifies the scattering outcomes.
The main tool for our analysis was the CQC formalism that enabled us to conveniently use classical 
equations of motion for the quantum fields. We treated quantum backreaction on the classical background 
by employing the semiclassical approximation. 
The main result of this work is summarized in Fig.~\ref{fig:resonance} where we see changes
in the fractal structure of the scattering as the interaction strength ($\xi$) between the kinks and
the quantum field increases. For $\xi \gtrsim 0.1$, the fractal structure disappears and there is simply
annihilation for $v_{\rm in} \lesssim 0.37$ and reflection for $v_{\rm in} \gtrsim 0.37$.

In future we hope to return to examining the effect of quantum fluctuations of the background scalar 
field itself on the scattering outcome, as mentioned in the Introduction (see Eq.~\eqref{phisplit}). 
To obtain a correspondence between our two field case with $\phi$ and $\psi$ and the
single field case in \eqref{phisplit}, we can add the equations of motion for $\phi$ and $\psi$
and compare to the equation of motion obtained using \eqref{phisplit}. The two equations
agree for the special choice of parameters $\mu^2 = -\lambda \eta^2$ and $\xi = 3 \lambda$.
Our technique fails for these values because then the $\psi$ field has a zero mode
(corresponding to the translation mode of the kink).
It still might be possible to consider the two field case in the limit that $\mu^2 \to -\lambda \eta^2$ 
and $\xi \to 3 \lambda$ and obtain a reasonable approximation to the single field case.
Alternatively, perhaps recent progress in dealing with systems containing classical and quantum components
will be helpful~\cite{Bojowald:2020emy,Oppenheim:2023mox,Oppenheim:2023izn}.
\acknowledgements
We thank Omer Faruk Albayrak, Fabio van Dissel and George Zahariade for useful discussions
and Evangelos I. Sfakianakis for comments. 
M.\,M. is supported by NSF Grant No. AST-2108466. M.\,M. also acknowledges support from the 
Institute for Gravitation and the Cosmos (IGC) Postdoctoral Fellowship. M.\,M. wishes to thank the 
Yukawa Institute for Theoretical Physics (YITP), Kyoto University for hospitality where a major part 
of this work was done. T.\,V. is supported by
the U.S. Department of Energy, Office of High Energy Physics, under Award No.~DE-SC0019470.
\appendix 
\section{Numerical methods}
\label{appsec:numerics}
We use the \emph{Verlet} method to solve the coupled differential equations for this work. In particular, we use the position Verlet method as follows,
\ba
x_{n+1/2} &=& x_n + \frac{1}{2} dt\ \dot{x}_n\,, \nn \\
\dot{x}_{n+1} &=& \dot{x}_n + dt\ \ddot{x}(x_{n+1/2})\,, \\
x_{n+1} &=& x_{n+1/2} + \frac{1}{2} dt\ \dot{x}_{n+1}\,. \nn
\ea
Although fairly simple, this method is efficient, fast, and gives us accurate results for the current problem. 
More advanced numerical techniques may be used but given the $\mathcal{O}(N^2)$ complexity of the 
problem, the requirements of fine spatial and temporal resolution to capture the details of the dynamics, 
and the necessity to evolve the system for long durations, the current numerical scheme is ideal. 
We have also verified our benchmark results using a more advanced numerical technique - the explicit 
Crank-Nicholson method with two iterations, and found no significant difference in the evolution and 
observable quantities. All the results presented are simulated on a circular lattice of physical length $L=100$ 
which is sampled using $N=500$ points. This implies a spatial resolution $a = L/N = 0.2$. We use a temporal 
spacing proportional to $a$, $dt = a/5$. A smaller $a$ would indeed improve the resolution and help us to 
capture the dynamics even better, but with the current lattice spacing we have a total energy non-conservation 
due to numerical errors, over the \emph{entire duration of time evolution} of the order of $0.2\%$. We have also checked that our results are fairly independent of the UV 
($a=L/N$) and IR ($L$) cutoffs. We start the evolution at the initial time, $t_0 = -100$. The collisions happen 
around $t \sim 0$. The periodic boundary conditions make it necessary to ensure that the finite lattice size 
does not significantly interfere with our results. Owing to our use of periodic boundary conditions, the radiation propagating outwards from the center eventually comes back. Hence, we only evolve the dynamics for one light 
crossing time, $t=L=100$, such that our results are not affected by radiation that re-enters the collision region.

For all the results form here on we assume, $\lambda=1$ and $\eta=1$. We choose $\mu = 0.1\ m$. 
This in the regime where $\mu < m$, which is required since $m$ is the mass of the classical background 
which we expect to be larger than the mass of the quantum field. The two main parameters that we vary 
are the initial velocity, $v_{\rm in}$, and the interaction strength, $\xi$. The initial incoming velocity of 
the kink-antikink configuration $v_{\rm in}$ is varied 
between $0.15$ and $0.4$. We do not consider lower velocities since for such low velocities the final state is 
fixed to be a bound state and no resonance structures are present, and in this work we are interested in 
studying quantum modifications to the resonance structures. We do not consider $v_{\rm in} > 0.4$ since
kinks with high initial velocities get reflected.
We vary $\xi$ in the range $[0,0.1]$.
\section{Analysis of the quality of numerics}
\label{appsec:quality}
\begin{figure*}
\subfloat[(a)] {\includegraphics[width=0.49\textwidth]{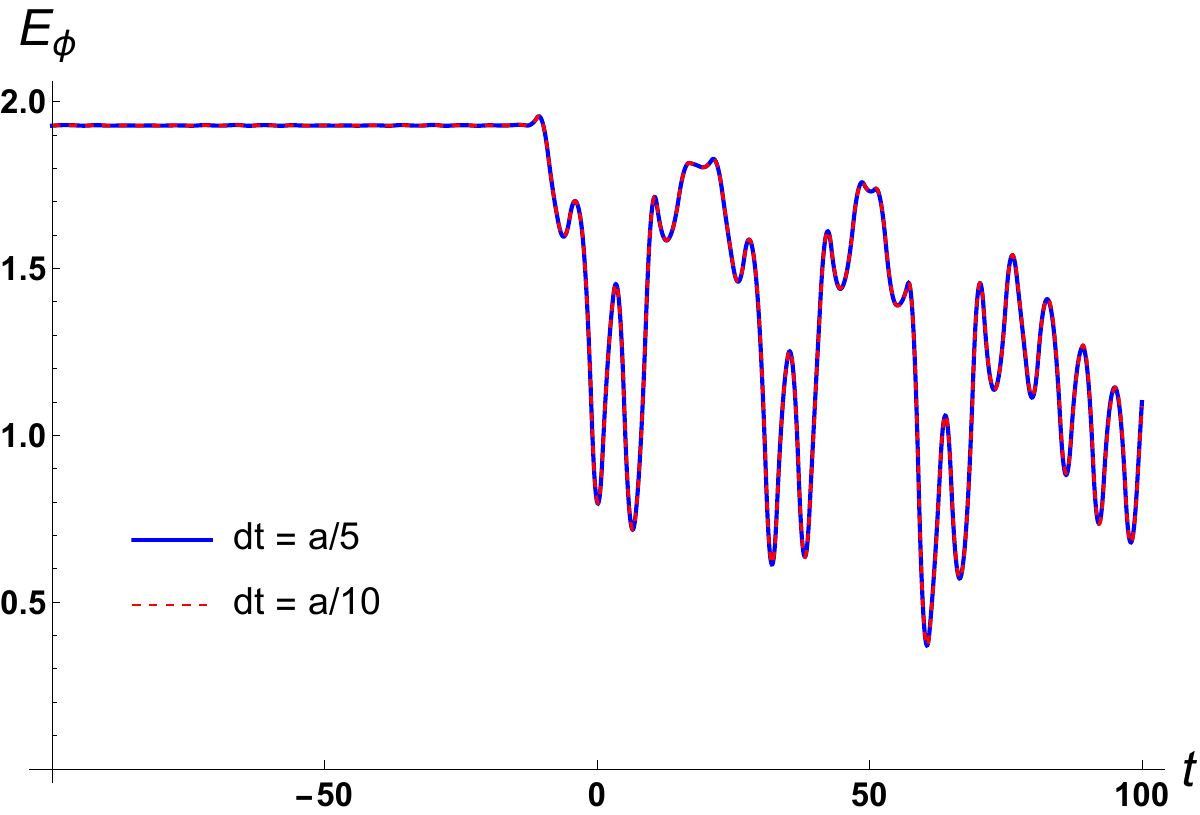}\label{fig:ephi_check}}\hfill
\subfloat[(b)] {\includegraphics[width=0.49\textwidth]{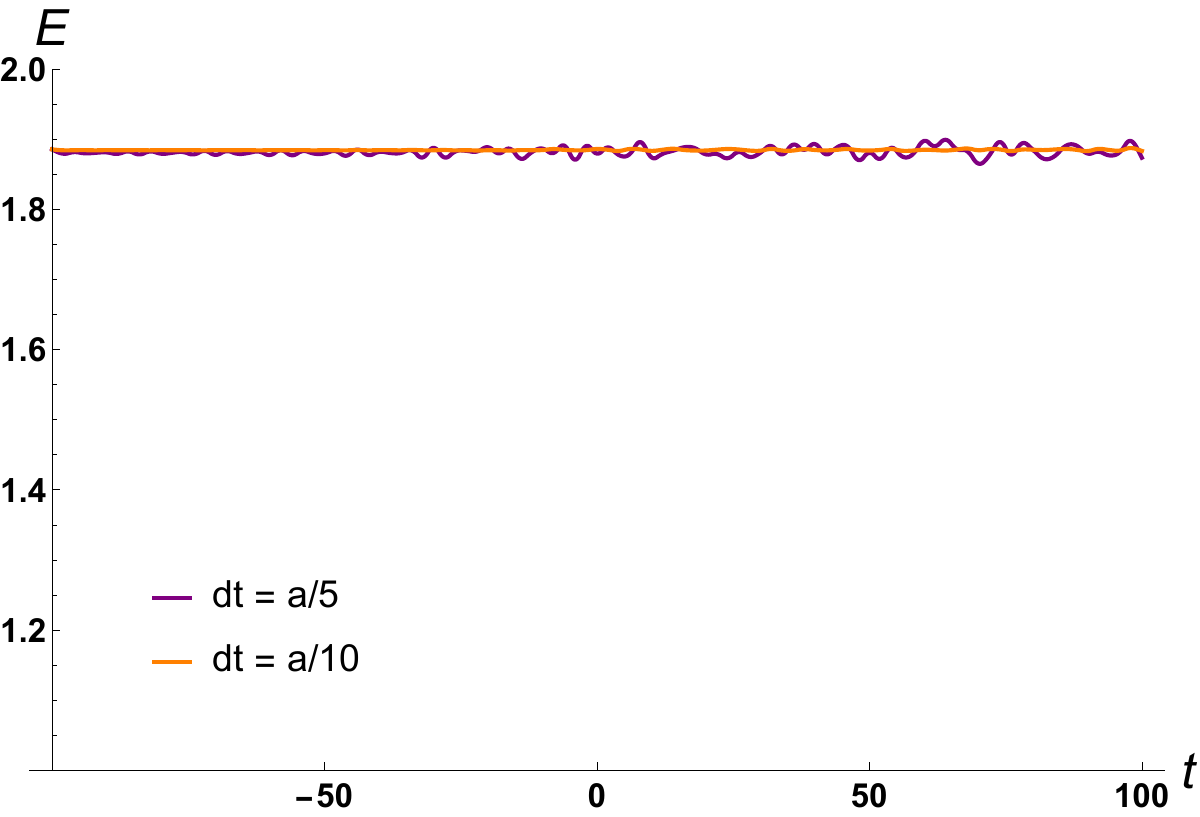}\label{fig:toten_check}} \hfill
\caption{\label{fig:dt_checks} (a) Time-evolution of the background energy ($E_\phi$) for different values of $dt$. The two curves overlap and are not distinctly visible.
(b) Time-evolution of the renormalized total energy $E$ for different values of $dt$. The plots illustrate the independence of our results to the choice of time step. The other parameters considered are: $v_{\rm in} = 0.21$, $\xi = 0.5$, $\mu = 0.1$, $\lambda=1$, $\eta = 1$, $t_0=-100$, $L=100$, and $N=500$. The collision happens at $t=0$.
}
\end{figure*}
The choice of time-step ($dt$) should not affect our results in any considerable way. This is shown in Fig.~\ref{fig:dt_checks}. The energy in the background does not show any difference on using a smaller time step ($dt=a/10$, dashed red curve) than what is used for this work ($dt = a/5$, solid blue curve). The total energy non-conservation for the time step we use ($dt=a/5$, solid purple curve) is of the order of $\sim 0.2\%$ over the \emph{entire} time of evolution. This is more than sufficient for the current work. The accuracy of energy conservation increases even more if we take a smaller $dt$ ($=a/10$, solid orange curve), which is to be expected.
\begin{figure}
\includegraphics[width=0.49\textwidth]{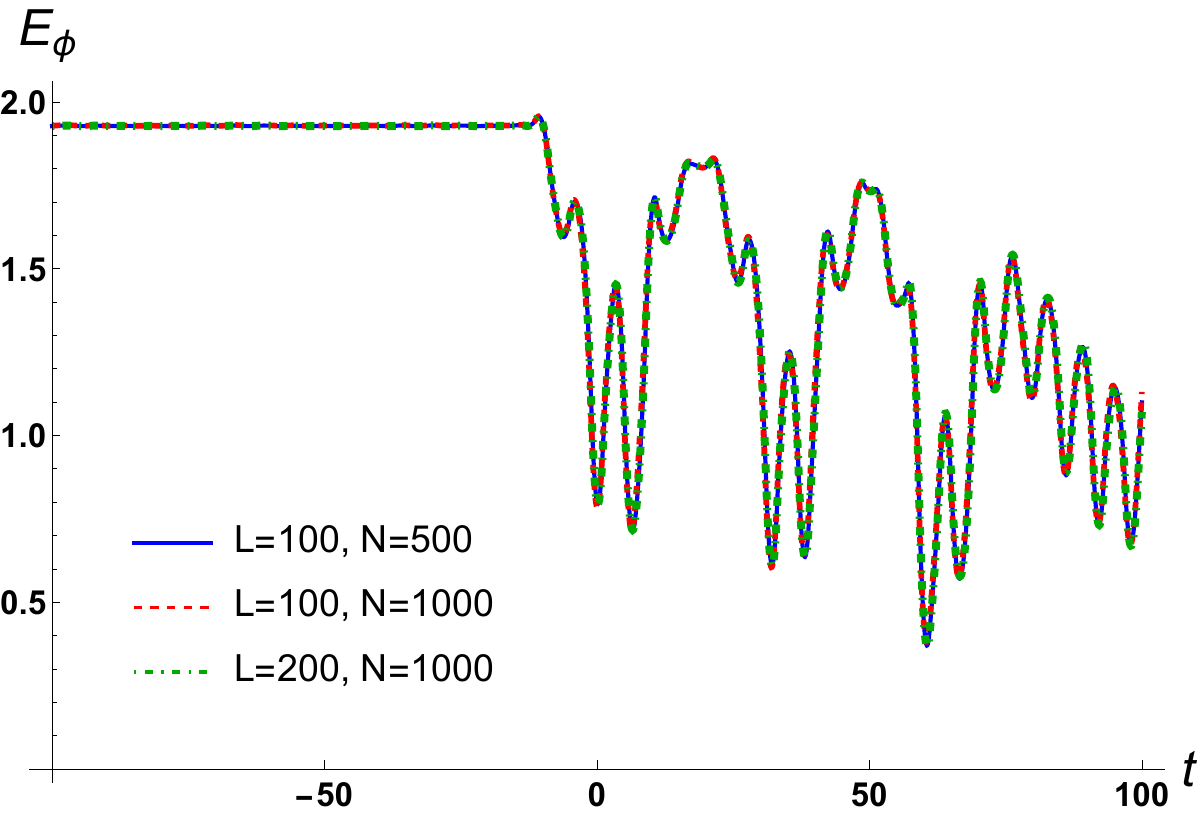}
\caption{\label{fig:ren_uv_ir} Time-evolution of the background energy ($E_\phi$) for different values of $L$ and $N$ 
to illustrate the UV- and IR-dependence of our results. The other parameters considered are: $v_{\rm in} = 0.21$, 
$\xi = 0.5$, $\mu = 0.1$, $\lambda=1$, $\eta = 1$, and $t_0=-100$. The collision happens at $t=0$. 
Note, lattice spacing $a=L/N$ and $dt = a/5$.
}
\end{figure}

One of the main physical observables we are interested in is the energy  ($E_\phi$)  in the background field $\phi$ 
and the total energy ($E$) (as discussed in Sec.~\ref{subsec:energy}). As one might expect, these observables 
should not depend on the discretization scales - the spacing of the lattice ($a$) and the physical size of the lattice 
($L$). 
In Fig.~\ref{fig:ren_uv_ir} we show such is the case, where decreasing the lattice spacing 
or increasing the physical size of the box has no effect on $E_\phi$. 
\bibstyle{aps}
\bibliography{refs}

\end{document}